\documentclass[onecolumn,floatfix,superscriptaddress,a4paper,
               showpacs,showkeys,nofootinbib,preprint]{revtex4}
\usepackage{epsfig}
\usepackage{latexsym}
\usepackage{xspace}
\usepackage{hyperref}
\usepackage[latin2]{inputenc}
\usepackage{indentfirst}
\usepackage{enumerate}

\usepackage{amsmath}
\usepackage{amssymb}
\usepackage[english]{babel}
\usepackage{url}

\newcommand{\eq}[1]{\begin{align} #1 \end{align}}

\begin{document}

\title{On Normalization of Strongly Intensive Quantities\\
       }

\author{M. Gazdzicki}
 \affiliation{Geothe-University Frankfurt am Main, Germany}
 \affiliation{Jan Kochanowski University, Kielce, Poland}

 \author{M. I. Gorenstein}
 \affiliation{Bogolyubov Institute for Theoretical Physics, Kiev, Ukraine}
 \affiliation{Frankfurt Institute for Advanced Studies, Frankfurt, Germany}

\author{M. Mackowiak-Pawlowska}
 \affiliation{Geothe-University Frankfurt am Main, Germany}
 \affiliation{Warsaw University of Technology, Warsaw, Poland}
\begin{abstract}
A special normalization is proposed for strongly intensive quantities used in
the study of event-by-event
fluctuations in high energy collisions. It ensures that these
measures are dimensionless and yields a common scale required for
a quantitative comparison of fluctuations of different, in general
dimensional, extensive quantities. Namely, the properly normalized
strongly intensive quantities assume the value one for
fluctuations given by the model of independent particle production
and zero in the absence of event-by-event fluctuations.
\end{abstract}

\pacs{12.40.-y, 12.40.Ee}

\keywords{}

\maketitle

\section{Introduction}\label{intr}
Intensive quantities are defined within the grand canonical
ensemble of statistical mechanics. They depend on temperature
and chemical potential(s), but they are independent of the system
volume.
Strongly intensive quantities~\cite{GG:2011} are, in addition,
independent of volume fluctuations. They were suggested for and
are used in studies of event-by-event fluctuations of hadron
production in nucleus--nucleus collisions at high energies. This
is because in these collisions, the volume of created states
varies from collision to collision, and is difficult or even
impossible to measure.

Strongly intensive quantities are defined using two arbitrary,
{\it extensive state quantities} $A$ and $B$. Here, we call $A$
and $B$ {\it extensive} when the first moments of their
distributions for the ensemble of possible states is proportional
to volume. They are referred to as {\it state quantities} as they
characterize the states of the considered system, e.g., final
states (or equivalently events) of nucleus--nucleus collisions or
micro-states of the grand canonical ensemble. For example, $A$ and
$B$ may stand for multiplicities of pions and kaons in a
particular state, respectively.

The simplest family of strongly intensive quantities is given by the
ratio of the first moments (i.e., average values) of $A$ and $B$:
\eq{\label{Ratio-AB} R[A,B]
 ~=~ \frac {\langle A\rangle} {\langle B\rangle}~,
}
where averaging $\langle \ldots \rangle$ is performed over the
ensemble of considered states.

There are two families of strongly intensive quantities which
depend on the second and first moments of $A$ and $B$ and thus
allow the study of state-by-state fluctuations~\cite{GG:2011}. These
are:
 \eq{\label{Delta-AB}
 \Delta[A,B]
 ~&=~ \frac{1}{C_{\Delta}} \Big[ \langle B\rangle\,
      \omega[A] ~-~\langle A\rangle\, \omega[B] \Big]~,
 \\
  \Sigma[A,B]
 ~&=~ \frac{1}{C_{\Sigma}}\Big[
      \langle B\rangle\,\omega[A] ~+~\langle A\rangle\, \omega[B] ~-~2\left(
      \langle AB \rangle -\langle A\rangle\langle
      B\rangle\right)\Big]~,
 \label{Sigma-AB}
 }
where
 \eq{\label{omega-AB}
 \omega[A] ~\equiv~ \frac{\langle A^2\rangle ~-~ \langle A\rangle^2}{\langle
 A\rangle}~,~~~~
 \omega[B] ~\equiv~ \frac{\langle B^2\rangle ~-~ \langle B\rangle^2}{\langle
 B\rangle}~
 }
are scaled variances of $A$ and $B$.
The normalization factors  $C_{\Delta}$ and
$C_{\Sigma}$ are required to be
proportional to the first moment of any extensive quantity.

It is important to stress that  $\Delta[A,B]$ and $\Sigma[A,B]$
are independent of system size fluctuations not only for the grand
canonical ensemble of states. They are also independent of the
average number of sources and source number fluctuations in the model of
independent particle sources, for example, in the wounded nucleon
model~\cite{WNM}.

Strongly intensive quantities for fluctuations have a long
history. The first quantity of this type, introduced in 1992, was
the so-called  $\Phi$ measure of fluctuations~\cite{GM:1992}.
According to the current classification the $\Phi$ measure belongs
to the $\Sigma$ family~\cite{GG:2011}. It is defined as the
difference of the quantity calculated for a studied ensemble
(e.g., central Pb+Pb collisions) and its value obtained within an
independent particle model (IPM) which preserves basic features of
the  ensemble. Thus, by construction, $\Phi = 0$ if the studied
ensemble satisfies the assumptions of the IPM. In general, $\Phi$
is a dimensional quantity and it does not assume a characteristic
value for the case of non-fluctuating $A$ and $B$. The latter
properties  were clearly disturbing in numerous applications of
$\Phi$ when attempting to characterize fluctuations in
experimental data~\cite{Phi_data} and models~\cite{Phi_models}.

In this paper we propose a specific choice of the $C_{\Delta}$ and
$C_{\Sigma}$ normalization factors which makes the quantities
$\Delta[A,B]$ and $\Sigma[A,B]$ dimensionless and  leads to
$\Delta[A,B] = \Sigma[A,B] = 1$ in the IPM. Moreover, from the
definition of $\Delta[A,B]$ and $\Sigma[A,B]$ it follows that
$\Delta[A,B] = \Sigma[A,B] = 0$ in the case of absence of
fluctuations of $A$ and $B$, i.e., for $ \omega[A] = \omega[B] =
\langle AB \rangle -\langle A\rangle \langle B \rangle = 0 $.
Thus the proposed normalization of $\Delta[A,B]$ and $\Sigma[A,B]$
leads to a common scale on which these fluctuation measures
calculated for different state quantities $A$ and $B$ can be compared.

The paper is organized as follows. In Section~\ref{IPM} we
introduce an independent particle model within which we calculate
the $\Delta[A,B]$ and $\Sigma[A,B]$ quantities. The calculation
details  are given in Appendix~\ref{IPM-APP}.
Appendix~\ref{AB-examples} gives explicit expressions of
$\Delta[A,B]$ and $\Sigma[A,B]$ for three choices of the quantities
$A$ and $B$.  Specific models which share the properties of the
IPM are discussed in Section~\ref{Models_examples}.
Section~\ref{procedure} presents the proposal for the
normalization of $\Delta[A,B]$ and $\Sigma[A,B]$,  discusses the
procedure to calculate them from a given ensemble of states and
provides an illustration by numerical examples. A summary in Section
\ref{sum} closes the article.

\section{$\Sigma[A,B]$ and $\Delta[A,B]$ in an Independent Particle Model}\label{IPM}
The independent particle model assumes that:
\begin{enumerate}[(i)]
\setlength{\itemsep}{1pt}
\item the state quantities $A$ and $B$ (e.g., of a micro-state of the grand canonical
ensemble, or of a final state of nucleus-nucleus collisions)
can be expressed as
\eq{\label{ind-part}
A~=~\alpha_1~+\alpha_2~+\ldots~+ \alpha_{N}~,~~~~~B~=~\beta_1~+\beta_2~+\ldots~+
\beta_{N}~,
}
where  $\alpha_j$ and $\beta_j$ denote single particle
contributions to $A$ and $B$, respectively, and $N$ is the number
of particles;
\item inter-particle correlations are absent, i.e. the
probability of any multi-particle state is the product of
probability distributions $P(\alpha_j,\beta_j)$ of single-particle
states, and these probability distributions are the same for all
$j=1,\ldots, N$ and independent of $N$~:
\eq{ \label{dist-alpha} P_N (\alpha_1,\beta_1,\alpha_2,\beta_2,
\dots, \alpha_N,\beta_N) = {\cal P}(N)\times
P(\alpha_1,\beta_1)\times  P(\alpha_2,\beta_2)\times \cdots \times
P(\alpha_N,\beta_N)~, }
where ${\cal P}(N)$ is an arbitrary multiplicity distribution of
particles.
\end{enumerate}

It is easy to show (see Appendix~\ref{IPM-APP}) that within the IPM
the average values of the first and second moments of $A$ and $B$
are equal to:
 \eq{\label{A}
 &\langle A \rangle~=~\overline{\alpha} ~\langle N\rangle~,~~~~
 \langle A^2 \rangle~=~\overline{\alpha^2}~\langle N \rangle ~+~
 \overline{\alpha}^2
 ~\left[\langle N^2\rangle ~-~\langle N\rangle\right]~,\\
&\langle B \rangle~=~\overline{\beta}~\langle N\rangle~,~~~~
 \langle B^2 \rangle~=~\overline{\beta^2}~\langle N \rangle ~+~
 \overline{\beta}^2
 ~\left[\langle N^2\rangle ~-~\langle N\rangle\right]~,\label{B} \\
 &\langle AB \rangle~=~
\overline{\alpha\,\beta}~\langle N \rangle ~+~
 \overline{\alpha}\,\cdot\,\overline{\beta}
 ~\left[\langle N^2\rangle ~-~\langle
 N\rangle\right]~.\label{AB}
}
The values of $\langle A \rangle$ and $\langle B \rangle$ are
proportional to the average number of particles $\langle N
\rangle$ and, thus, to the average size  of the system. These
quantities are extensive.  The quantities
$\overline{\alpha}$, ~$\overline{\beta}$ and
$\overline{\alpha^2}$, ~$\overline{\beta^2}$,
~$\overline{\alpha\,\beta}$ are the first and second moments of
the single-particle distribution $P(\alpha,\beta)$. Within the IPM
they are independent of $\langle N \rangle$ and play the role of
intensive quantities.

Using Eq.~(\ref{A}) the scaled variance $\omega[A]$ which
describes the state-by-state fluctuations of $A$ can be expressed as:
 \eq{\label{omega-A}
  \omega[A]~\equiv~ \frac{\langle A^2\rangle
~-~\langle A \rangle ^2}{\langle
A\rangle}~=~\frac{\overline{\alpha^2}~-~\overline{\alpha}^2}
{\overline{\alpha}}~+~\overline{\alpha}~\frac{\langle N^2\rangle
~-~\langle N\rangle^2}{\langle N\rangle}
~\equiv~\omega[\alpha]~+~\overline{\alpha}~ \omega[N]~,
}
where $\omega[\alpha]$ is the scaled variance of the
single-particle quantity $\alpha$, and $\omega[N]$ is the scaled
variance of $N$.
A similar expression follows from Eq.~(\ref{B}) for the scaled
variance $\omega[B]$. The scaled variances $\omega[A]$ and
$\omega[B]$ depend on the fluctuations of
the particle number via $\omega[N]$. Therefore, $\omega[A]$ and
$\omega[B]$ are not strongly intensive quantities.

From Eqs.~(\ref{A}-\ref{AB}) one obtains expressions for
$\Delta[A,B]$ and $\Sigma[A,B]$, namely:
\eq{
\Delta[A,B]~&=~ \frac{\langle N \rangle}{C_{\Delta}}~
\Big[~\overline{\beta}~ \omega[\alpha] ~-~\overline{\alpha}
~\omega[\beta]~\Big]~, \label{IPM-D}\\
\Sigma[A,B]~&=\frac{\langle N
\rangle}{C_{\Sigma}}~\Big[~\overline{\beta}~\omega[\alpha]~+~\overline{\alpha}~
\omega[\beta] ~-~2\left(~\overline{\alpha\, \beta}
-\overline{\alpha}\cdot \overline{\beta}~\right)~\Big]~.
\label{IPM-S}
}
Thus, the requirement that
\eq{\label{DS=1}
\Delta[A,B]~ = ~\Sigma[A,B]~ =~ 1~,
}
within the IPM leads to:
\eq{
C_{\Delta}~&=~\langle N \rangle~ \Big[~\overline{\beta}~
\omega[\alpha] ~-~\overline{\alpha}
~\omega[\beta]~\Big]~, \label{C-D}\\
C_{\Sigma}~&=~\langle N
\rangle~\Big[~\overline{\beta}~\omega[\alpha]~+~\overline{\alpha}~
\omega[\beta] ~-~2\left(~\overline{\alpha\, \beta}
-\overline{\alpha}\cdot \overline{\beta}~\right)~\Big]~.
\label{C-S}
}

Two comments are in order here. First,
\mbox{Eqs.~(\ref{A}-\ref{AB})} have the same structure as
\mbox{Eqs.~(2-4)} of Ref.~\cite{GG:2011} obtained within the model of
independent sources. The only difference is that the number of
sources $N_S$ in the model of independent sources is replaced by
the number of particles $N$ in the IPM. Each source can produce
many particles, and the number of these particles varies from
source to source and from event to event. Besides, the physical
quantities for particles emitted from the same source may be
correlated. Therefore, in general, the model of independent
sources does not satisfy the assumptions of the IPM. Nevertheless,
the formal similarity between the two models can be exploited and
gives the following rule of one to one correspondence: all results
for the IPM can be found from the expressions obtained within the
model of independent sources, assuming artificially that each
source always produces exactly one particle.  Second, only the
first and second moments of two extensive quantities $A$ and $B$
are required in order to define the strongly intensive quantities
$\Delta$ and $\Sigma$. However, in order to calculate the proposed
normalization factors $C_{\Sigma}$ and $C_{\Delta}$ additional
information is needed, namely the first and second moments of
single-particle contributions to $A$ and $B$ as well as the mean
number of particles.
Note that in special cases the factors $C_{\Sigma}$ and
$C_{\Delta}$ may assume the value zero and thus the proposed
normalization is not possible.

Explicit expressions for Eqs.~(\ref{C-D}, \ref{C-S}) for three
choices of $A$ and $B$ are given in Appendix~\ref{AB-examples}.
The first two cases correspond to the study of ''transverse momentum''
and ''chemical'' fluctuations. The third choice is the most general.

\section{Examples of Independent Particle Models }\label{Models_examples}
In this section  two specific models which satisfy the IPM
assumptions, i.e., Eqs.~(\ref{ind-part},~\ref{dist-alpha}), are
presented and discussed.
\subsection{Grand Canonical Ensemble}\label{GCE}
The most popular model which satisfies the IPM assumptions  is the
ideal Boltzmann multi-component gas in the grand canonical
ensemble  formulation. Here we refer to it as the IB-GCE. In the
IB-GCE the probability of any microscopic state is equal to the
product of probabilities of single-particle states. These
probabilities are independent of particle multiplicity. Thus, the
IB-GCE satisfies the assumption (\ref{dist-alpha}) of  the IPM.

The IB-GCE predicts a specific form of the multiplicity distribution
${\cal P}(N)$, namely the Poisson distribution and thus, $\omega[N]=1$.
Moreover, it also predicts the specific form of the single-particle
probability in momentum space, namely the Boltzmann
distribution:
\eq{\label{Boltz}
f_B({\bf p})~=~ C\,\exp\left(-\,\frac{\sqrt{{\bf
p}^2+m^2}}{T}\right)~,
}
where ${\bf p}$ and $m$ are particle momentum and mass,
respectively, $T$ is the system temperature, and $C=\left[\int d^3p
\, f_B({\bf p})\right]^{-1}$ is the normalization constant.

Note that by introducing quantum statistics one destroys the
correspondence between the GCE and the IPM. This is because of
(anti-)correlation between particles in the same quantum state for
the (Fermi) Bose ideal gas. Moreover, correlations between
particles are introduced if instead of resonances their decay
products are considered. Note that it is necessary to include the
strong decays of resonances in order to compare the GCE
predictions to experimental results.

\vspace{0.3cm} The correspondence between the IB-GCE and the IPM
remains valid even if the volume varies from micro-state
to micro-state\footnote{The statistical ensembles with volume
fluctuations were discussed in Ref.~\cite{volume}} but
local properties of the system, i.e., temperature and chemical
potentials are independent of the system volume.
Let volume fluctuations  be given by the probability density
function  $F(V)$.
The averaging over all micro-states includes
the averaging over the  micro-states with fixed volume
and the averaging over the
volume fluctuations.
The volume fluctuations broaden the ${\cal P}(N)$
distribution and increase its scaled variance:
\eq{\label{omega-j}
\omega[N] ~\equiv~\frac{\langle N^2\rangle~-~\langle
N\rangle^2}{\langle N\rangle } ~=~1~+~\frac{\langle
N\rangle}{\langle V\rangle}\cdot\frac{\overline{
V^2}~-~\overline{V}^2}{\overline{V}} ~,
}
where 
%
%
 $\overline{ V^k}~\equiv~\int dV~ F(V)~V^k$~ for $k=1,2$.
%
%
The first term on the right hand side  of Eq.~(\ref{omega-j})
corresponds to the particle number fluctuations in the IB-GCE at
a fixed volume $V$ (i.e., this is the scaled variance of the Poisson
distribution), and the second term is the contribution due to the
volume fluctuations.  Equation~(\ref{dist-alpha}) remains valid
in this example, therefore, the IB-GCE with arbitrary volume
fluctuations  satisfies the IPM assumptions.

\subsection{Mixed Event Model}
The Mixed Event Model is defined by the Monte Carlo procedure
frequently used by experimentalists in order to create a sample of
artificial events in which correlations and fluctuations present
in the original ensemble of events are partly removed. Then the
original and mixed events are analyzed in the same way and the
corresponding results are compared in order to extract the magnitude
of a signal of interest, which by construction should be present
in the original events and absent in the mixed events.
The mixed event procedure is in particular popular in studies
of resonance production, particle correlations due to quantum
statistics and event-by-event fluctuations, see for examples
Ref.~\cite{mixed}.

There are many variations of the Mixed Event Model. Here we
describe the one which in the limit of an infinite number of the original
and mixed events gives results identical to the IPM.

The procedure to create a mixed event which corresponds to the
given ensemble of original events consists of two steps, namely:
\begin{enumerate}[(i)]
\setlength{\itemsep}{1pt} \item a mixed event multiplicity, $N$,
is drawn from the set of multiplicities of all original events;
\item $N$ particles for the mixed event are drawn
randomly with replacement from the set of
all particles from all original events.
\end{enumerate}
Then the steps one and two are repeated to create the next mixed
event and the procedure is stopped when the desired number of
mixed events is reached.
In the limit of an infinite number of original events, the probability
to have two particles from the same original event in a single
mixed event is zero and thus particles in the mixed events are
uncorrelated. Therefore, in this limit the mixed event model
satisfies the IPM assumptions. Note, for an infinite number
of mixed events, the first moments of all extensive quantities and
all single-particle distributions of the original and mixed events
are identical.

\section{Normalization and determination of $\Delta[A,B]$ and $\Sigma[A,B]$
         }\label{procedure}
The strongly intensive quantities $\Delta[A,B]$ and $\Sigma[A,B]$
were introduced for the study of state-by-state fluctuations of any
extensive quantities $A$ and $B$ in a given ensemble of states.
For example, states may refer to data for nucleus--nucleus
collisions recorded by an experiment or generated within a
Monte-Carlo model.
In this section, we first explicitly introduce our special normalization
of $\Delta[A,B]$ and $\Sigma[A,B]$ for the ensemble of
states. Then, we discuss the procedure of their determination.
%
%

\begin{figure}[t]
\centering
\includegraphics[width=0.495\textwidth]{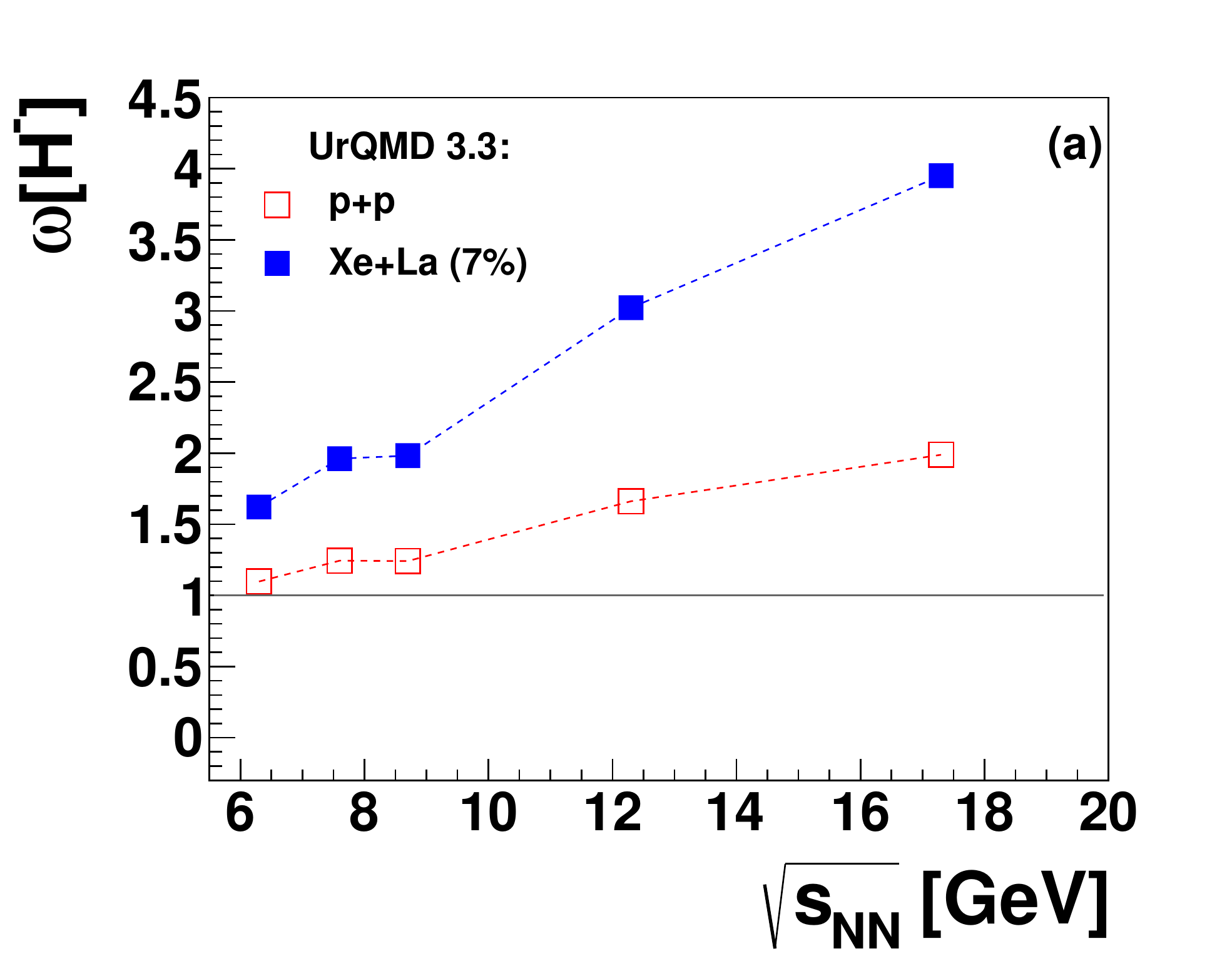}
\includegraphics[width=0.495\textwidth]{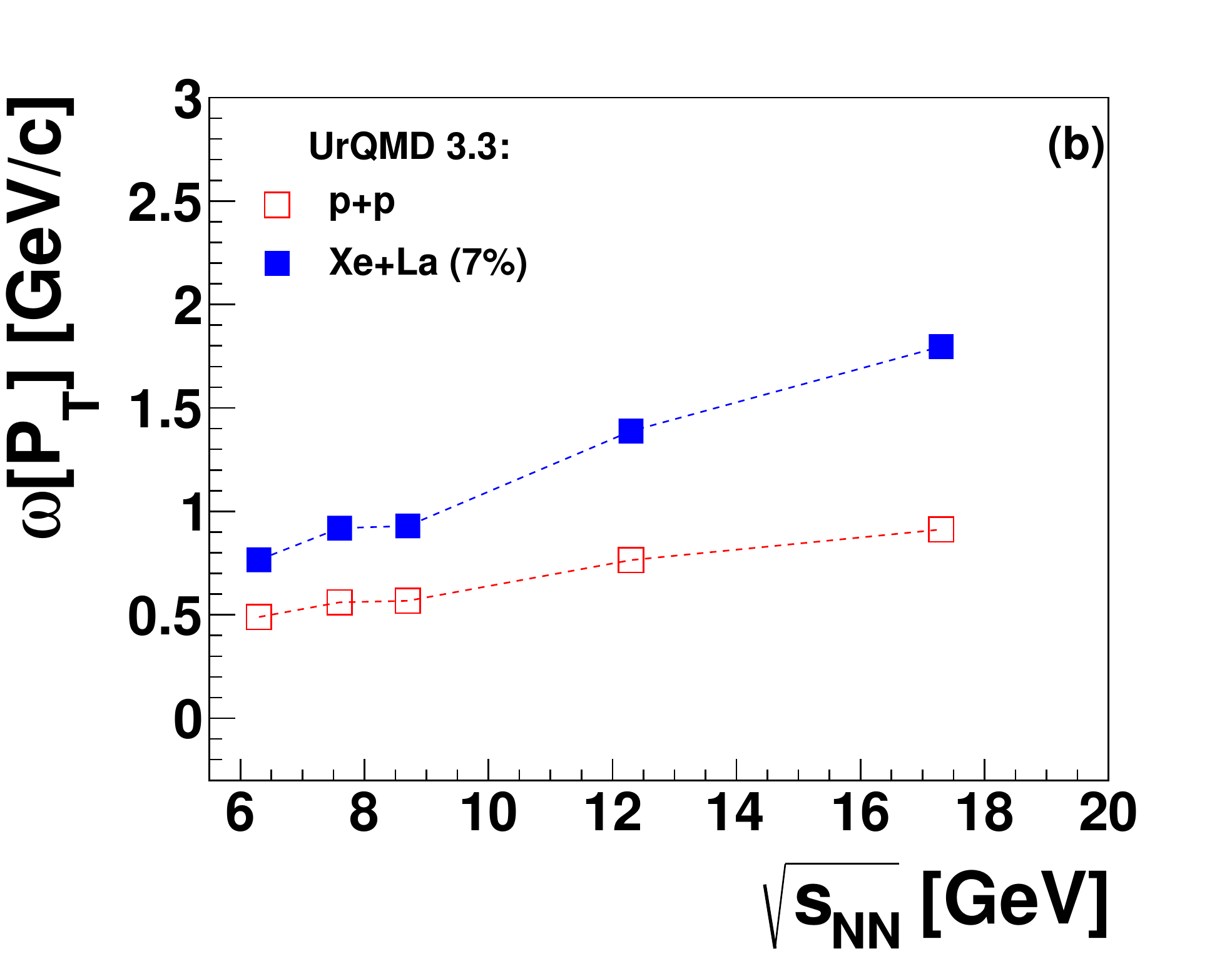}
\includegraphics[width=0.495\textwidth]{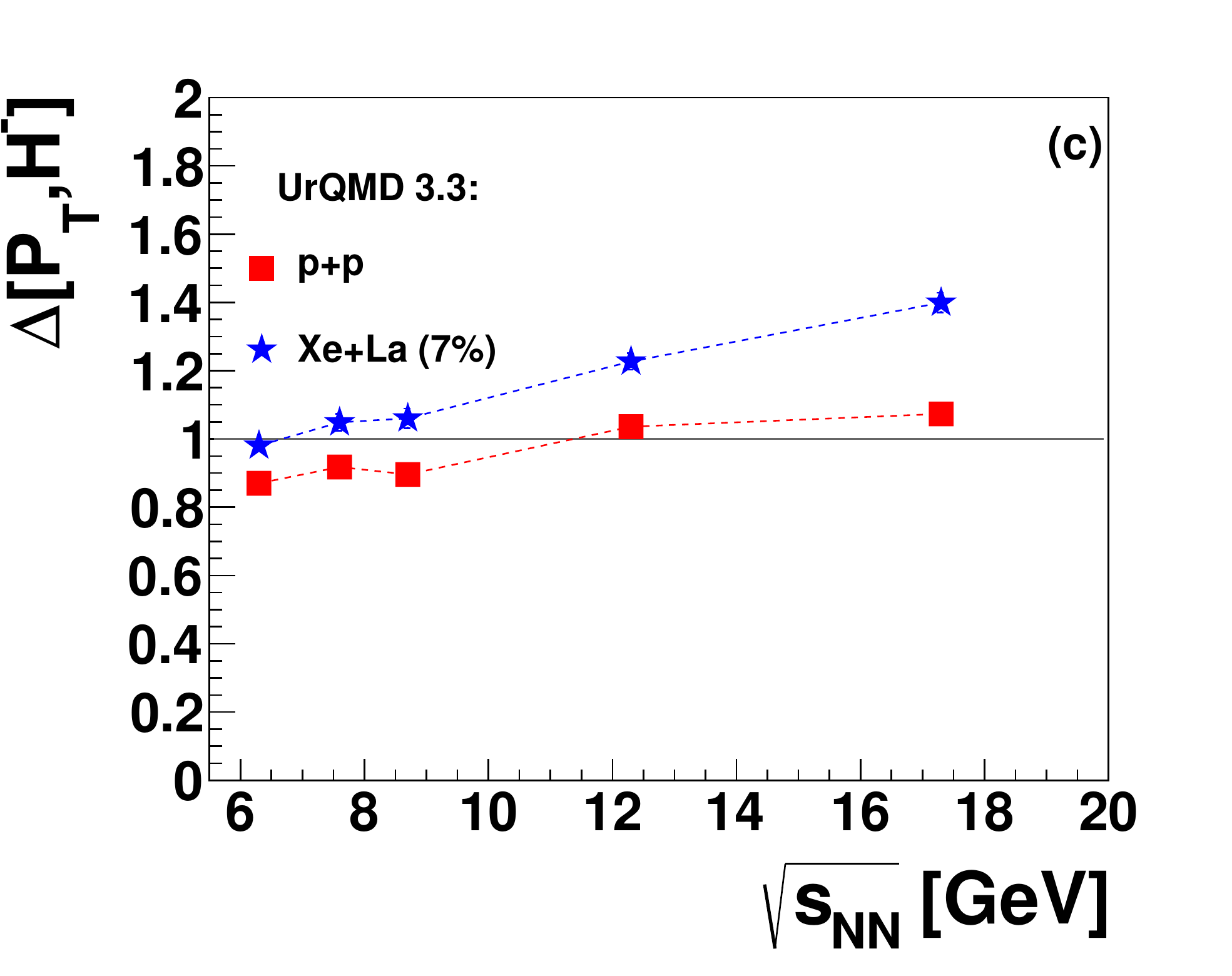}
\includegraphics[width=0.495\textwidth]{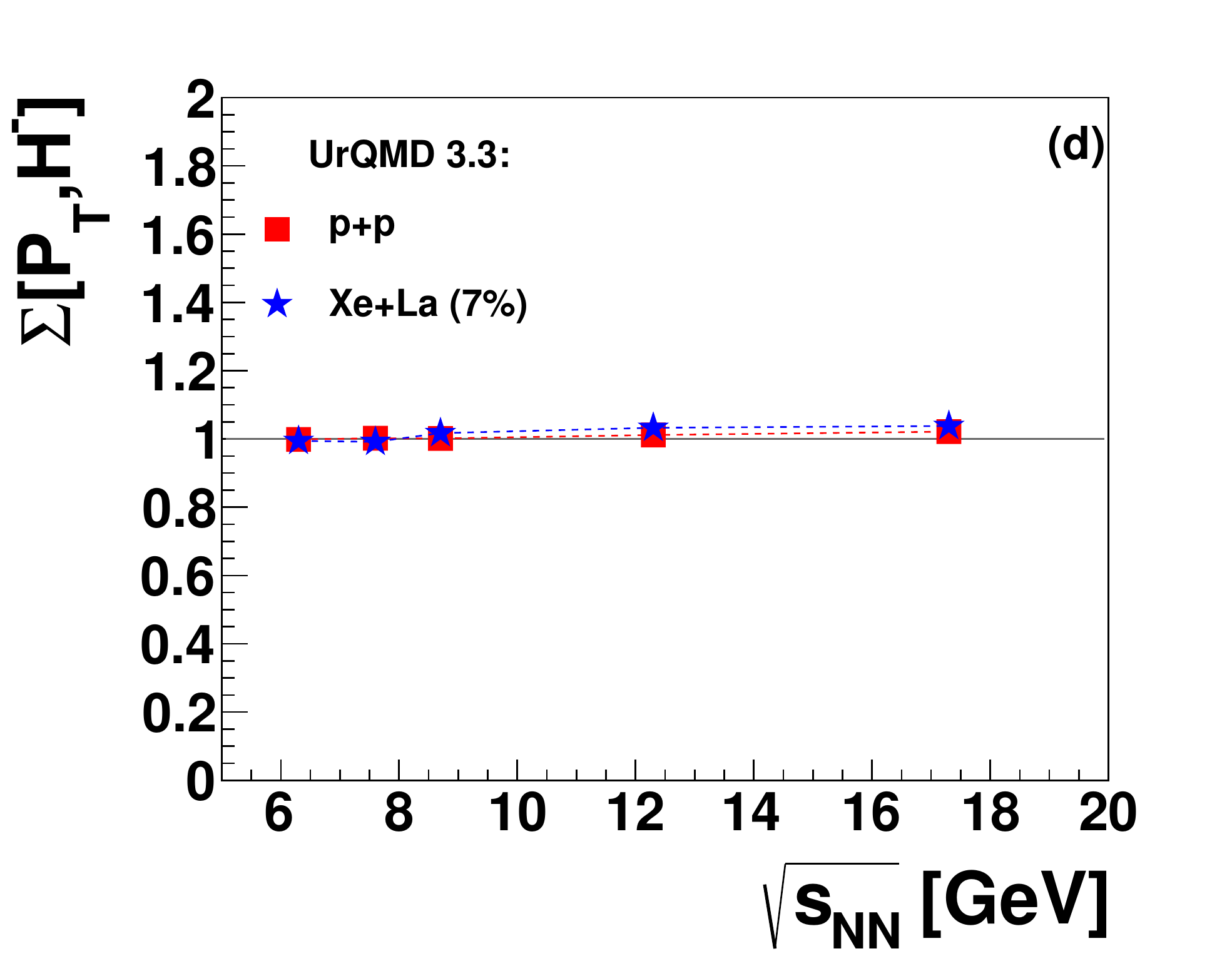}
\caption{ Fluctuation measures calculated within the
UrQMD model for negatively charged hadrons produced in inelastic p+p
interactions and the 7\% most central Xe+La collisions as functions of
collision energy in the CERN SPS energy range. The top plots show
intensive measures of fluctuations, namely the scaled variances of
(a) the negatively charged hadron multiplicity, $\omega[H^-]$, and (b)
the sum of magnitudes of
their transverse momenta, $\omega[P_T]$. The bottom plots show
the corresponding strongly intensive measures (c) $\Delta[P_T,H^-]$
and (d) $\Sigma[P_T,H^-]$.
Statistical uncertainties are smaller than the symbol size
and were calculated using the subsample method.
}
\label{fig:pt}
\end{figure}

Let the ensemble of states $\Omega$  and the extensive state
quantities $A$ and $B$ be given.
We propose to define the normalization factors $C_{\Sigma}$ and
$C_{\Delta}$ in Eqs.~(\ref{Delta-AB},~\ref{Sigma-AB}) such that
$\Delta[A,B]=\Sigma[A,B]=1$ in the IPM with the multiplicity distribution
${\cal P}(N)$ and the single-particle distribution $P(\alpha,\beta)$
identical to those of the ensemble $\Omega$. The IPM  which
corresponds to the ensemble $\Omega$ will be
denoted as the IPM-$\Omega$. The normalization factors $C_{\Delta}$
and $C_{\Sigma}$ calculated in the IPM-$\Omega$ are then given by
Eqs.~(\ref{C-D}, \ref{C-S}), where all entering quantities
should be calculated from the ensemble $\Omega$.
%
%

The procedure of calculating $\Delta[A,B]$ and $\Sigma[A,B]$ given
by Eqs.~(\ref{Delta-AB},~\ref{Sigma-AB}) with the normalization
factors defined by Eqs.~(\ref{C-D},~\ref{C-S}) consists of the following steps:
\begin{enumerate}[(i)]
\setlength{\itemsep}{1pt}
\item calculate the $\Omega$-ensemble state averages of the first and
second moments of extensive quantities $A$ and $B$;
\item calculate the first and second moments of single-particle
quantities, $\alpha$ and $\beta$, as well as the average number of particles
$\langle N\rangle$ entering
Eqs.~(\ref{C-D},~\ref{C-S});  the avareging is performed over the 
$\Omega$-ensemble states and particles;
\item calculate $\Delta[A,B]$ and $\Sigma[A,B]$ according to
Eqs.~(\ref{Delta-AB}) and (\ref{Sigma-AB}).
\end{enumerate}

The proposed procedure is illustrated by numerical results
obtained within the UrQMD model~\cite{UrQMD}.
Figure~\ref{fig:pt} shows the collision energy dependence of
different fluctuation measures discussed in this paper
in the CERN SPS energy range.
In this example, $A=P_T$ is the number of  negatively charged hadrons
and $B=H^-$ is their total transverse momentum (see Appendix \ref{examp-PT}).
The UrQMD simulations were performed
for inelastic p+p interactions and for the 7\% most central Xe+La collisions.
This choice of reactions is motivated by the experimental
program of the NA61/SHINE Collaboration~\cite{NA61} at the CERN SPS.
NA61/SHINE already reported
the first results on event-by-event fluctuations in p+p interactions~\cite{ppNA61},
and results for nucleus-nucleus (Be+Be, Ar+Ca, Xe+La)
collisions will become available within the next couple of years.
A comparison between experimental data and models is beyond the scope
of this paper.  
The top plots show intensive fluctuation measures, namely the scaled variance
of the negatively charged particle multiplicity distribution, $\omega[H^-]$, and
of the distribution of the sum of the magnitudes of their transverse momenta, $\omega[P_T]$.
The bottom plots show the corresponding strongly intensive measures
$\Delta[P_T,H^-]$  and $\Sigma[P_T,H^-]$ normalized as proposed in
this paper according to Eqs.~(\ref{C-D},~\ref{C-S}) with their
explicit form given in Eq.~(\ref{C-D-PT}).

The scaled variance of $H^-$ and $P_T$ is significantly larger
in central Xe+La collisions than in p+p interactions.
To a large extent this is due to fluctuations of the number
of nucleons which interacted (wounded nucleons),
see Ref.~\cite{Be:2012} for a detailed discussion of this issue.
The advantages of the $\Delta[P_T,H^-]$  and $\Sigma[P_T,H^-]$ quantities
are obvious from the results presented in the bottom plots.
First, they are not directly  sensitive to fluctuations of the collision
geometry (the number of wounded nucleons) in contrast to the scaled variance.
Thus, the remaining small differences between results for central Xe+La collisions
and p+p interactions are entirely due to deviations of the UrQMD  model
from the independent source model. 
Second, they are dimensionless and expressed
in units common for all energies and reactions as well as for
different choices of state
quantities $A$ and $B$.
Due to the particular normalization, proposed in this article,
they assume the value one
for the independent particle model and zero
in the absence of event-by-event fluctuations.

\section{Summary}\label{sum}

Strongly intensive quantities $\Delta[A,B]$ and $\Sigma[A,B]$ are
fluctuation measures which are independent of the system volume
and its fluctuations within the grand canonical ensemble of
statistical mechanics. Moreover, they are independent of the
number of wounded nucleons and its fluctuations within the Wounded
Nucleon Model. Strongly intensive quantities are expected to be
useful in studies of fluctuations in hadron production in
nucleus--nucleus collisions at high energies.
In this paper a special normalization of strongly intensive
quantities is proposed. It ensures that they are dimensionless
and yields a common scale enabling a quantitative comparison of
fluctuations of different extensive state quantities. With the
proposed normalization the discussed strongly intensive quantities
assume the value one for fluctuations given by the independent
particle model and zero in the absence of state-by-state
fluctuations.

The paper includes details of calculations, explicit formulas
for ''transverse momentum'' and  ''chemical'' fluctuations as well as
for the most general case of fluctuations of two extensive motional
quantities for partly overlapping sets of particles.
Moreover numerical examples are given using final states of
high energy collisions generated by the UrQMD model.

\vspace{0.3 cm}
\begin{acknowledgments} We are thankful to Viktor Begun, Katarzyna Grebieszkow,
Benjamin Messerly, Stanislaw Mrowczynski, Peter Seyboth and Anar
Rustamov for fruitful discussions and comments. This work was
supported by the Program of Fundamental Research of the Department
of Physics and Astronomy of NAS, Ukraine and the German Research
Foundation (grant DFG GA 1480/2-1).
\end{acknowledgments}
%

\appendix

\section{Calculation details of the IPM}
\label{IPM-APP}
In this Appendix details of the derivation of Eqs.~(\ref{A}-\ref{AB})
within the Independent Particle Model defined by
Eqs.~(\ref{ind-part},~\ref{dist-alpha}) are given.

The functions entering Eq.~(\ref{dist-alpha}) satisfy
the normalization conditions:
\eq{\label{norm}
\sum_N {\cal P}(N)~=~1~,~~~~~\int
d\alpha\,d\beta~P(\alpha,\beta)~=~1~.
}
The average values of $\alpha_j$ and $\alpha_j^2$ are:
\eq{\label{alpha-j}
&\langle \alpha_1\rangle=\langle \alpha_2\rangle=\ldots=\langle
\alpha_N\rangle~=~\int d\alpha d\beta\,\alpha\,P(\alpha,\beta)~
\equiv~\overline{\alpha}~,\\
&\langle \alpha_1^2\rangle=\langle
\alpha_2^2\rangle=\ldots=\langle \alpha_N^2\rangle~=~\int d\alpha
d\beta\,\alpha^2\,P(\alpha,\beta)~ \equiv~\overline{\alpha^2}~.
\label{alpha-2-j}
}
Similarly,
\eq{\label{beta-j}
&\langle \beta_1\rangle=\langle \beta_2\rangle=\ldots=\langle
\beta_N\rangle~=~\int d\alpha\, d\beta~\beta~P(\alpha,\beta)~
\equiv~\overline{\beta}~,\\
&\langle \beta_1^2\rangle=\langle \beta_2^2\rangle=\ldots=\langle
\beta_N^2\rangle~=~\int d\alpha\, d\beta ~\beta^2~P(\alpha,\beta)~
\equiv~\overline{\beta^2}~. \label{beta-2-j}
}
At $i\neq j$ one finds:
\eq{\label{i-neq-j}
\langle \alpha_i\alpha_j\rangle~=~\langle \alpha_i\rangle \langle
\alpha_j\rangle~=~\overline{\alpha}\cdot
\overline{\alpha}~=~\overline{\alpha}^2,~~~~
\langle \beta_i\beta_j\rangle~=~\langle \beta_i\rangle \langle
\beta_j\rangle~=~\overline{\beta}\cdot
\overline{\beta}~=~\overline{\beta}^2~.
}
The state averages of $A$ and $B$ are equal to:
\eq{\label{IPM-A}
& \langle A\rangle ~=~\langle~ \sum_{j=1}^N \alpha_j
~\rangle~=~\sum_N {\cal P}(N)\,\sum_{j=1}^N \langle \alpha_j
\rangle ~=~\sum_N {\cal P}(N)~ \overline{\alpha}\cdot
N~=~\overline{\alpha}\, \langle N\rangle~,\\
& \langle B\rangle ~=~\langle~ \sum_{j=1}^N \beta_j
~\rangle~=~\sum_N {\cal P}(N)\,\sum_{j=1}^N \langle \beta_j
\rangle ~=~\sum_N {\cal P}(N)~ \overline{\beta}\cdot
N~=~\overline{\beta}\, \langle N\rangle~.\label{IPM-B}
}
For the second moments of $A$ and $B$ one obtains:
\eq{\label{IPM-A2}
&\langle A^2\rangle ~=~\langle~
\left(\alpha_1+\alpha_2+\ldots+\alpha_N\right)^2~\rangle~ =~
\langle~ \sum_{j=1}^N \alpha_j^2 ~+~\sum_{1\neq i<j\leq N}
\alpha_i\alpha_j ~\rangle \nonumber\\
&=~ \sum_N {\cal P}(N)\,\left[\sum_{j=1}^N \langle \alpha_j^2
\rangle ~+\sum_{1\leq i\neq j\leq N} \langle \alpha_i\alpha_j
\rangle \right]~=~\overline{\alpha^2}\,\langle N\rangle
~+~\overline{\alpha}^2\,\left[\langle N^2\rangle -\langle
N\rangle\right]~,
 }
and similarly
\eq{\label{IPM-B2}
 \langle B^2\rangle ~=~\overline{\beta^2}\,\langle N\rangle
~+~\overline{\beta}^2\,\left[\langle N^2\rangle -\langle
N\rangle\right]~.
}
Finally, for $\langle AB\rangle$ one finds:
\eq{\label{IPM-AB}
&\langle AB\rangle~=~\langle~
\left(\alpha_1+\alpha_2+\ldots+\alpha_N \right)\times
\left(\beta_1+\beta_2+\ldots+\beta_N\right)~\rangle
\nonumber\\
&=~ \sum_N {\cal P}(N)\,\left[\sum_{j=1}^N \langle \alpha_j\beta_j
\rangle ~+~\sum_{1\leq i\neq j\leq N} \langle \alpha_i\beta_j
\rangle \right]~=~\overline{\alpha\beta}\,\langle N\rangle
~+~\overline{\alpha}\cdot \overline{\beta} \,\left[\langle
N^2\rangle -\langle N\rangle\right]~.
}
\section{Examples for three choices of $A$ and $B$}\label{AB-examples}
In this Appendix  explicit expressions for
$\Delta[A,B]$ and $\Sigma[A,B]$ and their normalization factors
$C_{\Sigma}$ and $C_{\Delta}$ calculated within the IPM are given
for  two popular choices for the extensive state quantities
$A$ and $B$ which correspond to the study of ''transverse momentum''
and ''chemical'' fluctuations. Finally the most general case is
considered, which corresponds to the selection of two extensive motional
quantities for partly overlapping sets of particles.
\subsection{''Transverse momentum'' fluctuations}
\label{examp-PT}
The first~\cite{GM:1992} and the most popular~\cite{Phi_data,Phi_models}
application of the $\Phi$ measure was the study of
transverse momentum fluctuations.
In the formalism, introduced in Ref.~\cite{GG:2011},
this corresponds to the following
choice of the extensive state quantities $A$ and $B$:
\eq{\label{PT}
& A \equiv P_T~=~p^{(1)}_T+p^{(2)}_T+\ldots+p^{(N)}_T~,
\\
& B \equiv N~=~~w^{(1)}+w^{(2)}+\ldots+w^{(N)}~,
}
where $p_T^{(j)}$ is the absolute value of the transverse momentum
of $j$$^{th}$ particle\footnote{Similarly one can consider  sums of
any other motional variables, e.g., particle energies, rapidities,
etc.}
and $w^{(j)}$ is the particle identity~\cite{Ga:1999} which
equals to one for all particles: $w^{(j)} = 1$.

Thus, for the single-particle quantities:
\eq{\label{PT_single}
\alpha~=~ p_T~,~~~~~~ \beta ~=~ w~=~1~,
}
one gets:
\eq{\label{ab-PT}
\overline{\alpha}= \overline{p_T}~, ~~~~\overline{\alpha^2}
=\overline{p_T^2}~,~~~~
\overline{\beta}=\overline{\beta^2}=\overline{w}=\overline{w^2}=1~,
~~~~\overline{\alpha \beta}
=\overline{p_T}~,
}
where $\overline{p_T}$ and $\overline{p_T^2}$ are the average
values of $p_T$ and $p_T^2$ calculated from the properly normalized
single-particle transverse momentum distribution\footnote{In high
energy physics single-particle distributions are called
{\it inclusive distributions}.}.
Consequently, Eqs.~(\ref{C-D}, \ref{C-S})  give:
\eq{\label{C-D-PT}
C_{\Delta}~=~C_{\Sigma}~=~ \langle N\rangle ~\cdot~
\frac{\overline{ p_T^2}~-~\overline{p_T}^2}
{\overline{p_T}}~\equiv~\langle N\rangle\,\cdot\,\omega[p_T]~.
}
As was already mentioned in Section~\ref{IPM}, only the first
and second moments of two extensive quantities $P_T$ and $N$ are
required to calculate the strongly intensive measures
$\Delta[P_T,N]$ and $\Sigma[P_T,N]$. However, in order to
calculate the proposed normalization factors $C_{\Delta}$ and
$C_{\Sigma}$ additional information may be necessary. In the
considered example, one also needs the
second moment of the single-particle $p_T$-distribution
$\overline{p_T^2}$ to calculate (\ref{C-D-PT}).

 Let us recall here that $\Sigma(P_T,N)$ is directly related
to the $\Phi_{p_T}$ measure of transverse momentum
fluctuations, for the explicit expression see Ref.~\cite{GG:2011}.
The only difference is in the scale used to quantify fluctuations
measured by both quantities.
Namley, the $\Phi_{p_T}$ measure
is defined as the difference of the event  quantity calculated for the
studied ensemble (e.g., central Pb+Pb collisions) and its value
obtained within the independent particle model.
Consequently, $\Phi_{p_T} = 0$ if the studied ensemble satisfies the
assumptions of the IPM. 
Moreover, $\Phi_{p_T}$ is a dimensional quantity
and does not assume a characteristic value for the case of
non-fluctuating $A$ and $B$.
These undesired properties of $\Phi_{p_T}$ are removed when fluctuations
are measured using $\Sigma(P_T,N)$ normalized as proposed in this article.

\subsection{''Chemical'' fluctuations}\label{examp-Kpi}
In the jargon of high energy nuclear physics ''chemical'' fluctuations
refer to fluctuations of particle-type composition of the system.
In order to
be specific let us  consider  relative fluctuations of the
number of charged pions $\pi\equiv \pi^++\pi^-$ and kaons $K\equiv
K^++K^-$:
\eq{\label{chemical}
&A~\equiv~ K~=~w^{(1)}_{K}+w^{(2)}_{K}+\ldots+w^{(N)}_{K}~,
~~~~
B~\equiv~\pi~=~w^{(1)}_{\pi}+w^{(2)}_{\pi}+\ldots+w^{(N)}_{\pi}~,
}
where $w^{(j)}_{\pi}$ and $w^{(j)}_{K}$ are the pion and kaon
identities of the $j$$^{th}$ particle\footnote{Similarly one can consider
sums of any other particle identities, e.g., negatively
charged particle, baryons, etc.}. Particle identities were
introduced first in Ref.~\cite{Ga:1999} and used in  the
study  of ''chemical''
fluctuations in terms of the
$\Phi$ measure~\cite{Ga:1999,Phi_data,Phi_models}.

In this example one defines the kaon $w_{K}^{(j)}$ and
pion $w_{\pi}^{(j)}$ identities as: $w_K^{(j)}=1$ and
$w_{\pi}^{(j)}=0$ if the $j$$^{th}$ particle is a kaon, and
$w_K^{(j)}=0$ and $w_{\pi}^{(j)}=1$ if the $j$$^{th}$ particle is a
pion.

For the single-particle quantities,
\eq{\label{chemical_single}
\alpha~=~ w_{K}~,~~~~~ \beta ~=~ w_{\pi}~,
}
one obtains:
\eq{\label{w-kpi}
\overline{w_K}=\overline{w_K^2}=\frac{\langle K\rangle}{\langle
N\rangle}~\equiv~k~,
~~~~~\overline{w_\pi}=\overline{w_\pi^2}=\frac{\langle
\pi\rangle}{\langle N\rangle}~=~1-k~,~~~~~
\overline{w_K\,w_\pi}=0~,
}
where $N=K+\pi$. Then from Eq.~(\ref{w-kpi}) follows:
\eq{\label{IPM-om-kpi}
& \omega\left[w_K\right]~\equiv~
\frac{\overline{w_K^2}~-~\overline{w_K}^2}{\overline{w_K}}~=~
1~-~k~, ~~~~~~\omega\left[w_\pi\right]~\equiv~
\frac{\overline{w_\pi^2}~-~\overline{w_\pi}^2}{\overline{w_\pi}}~=~
k~,\\
&\overline{w_Kw_\pi}~-~\overline{w_K}\,\cdot\,\overline{w_\pi}~=~-~k\cdot(1-k)~.
}
Therefore, Eqs.~(\ref{C-D}, \ref{C-S}) give:
\eq{\label{Kpi-CD}
C_{\Delta}~&=~
\langle N\rangle \cdot(1-2k) ~=~\langle \pi \rangle ~-~\langle K\rangle~ , \\
C_{\Sigma}~&=~ \langle N\rangle
~=~\langle \pi
\rangle~+~\langle K\rangle~. \label{Kpi-CS}
}
As seen from Eqs.~(\ref{Kpi-CD}, \ref{Kpi-CS}) the normalization
factors $C_{\Delta}$ and $C_{\Sigma}$  depend only on
the first moments of the  extensive state quantities $K$ and $\pi$.

However, in general more information is needed to
calculate  $C_{\Delta}$ and $C_{\Sigma}$.
As an illustration let us consider partly overlapping sets of
particles, e.g., the number of charged kaons $K=K^++K^-$ and all
negatively charged particles $H^-$.
The extensive state quantities $A$ and $B$ are:
\eq{\label{chemK}
& A \equiv K~=~w^{(1)}_{K}+w^{(2)}_{K}+\ldots+w^{(N)}_{K}~,
\\
& B \equiv
H^-~=~w^{(1)}_{-}+w^{(2)}_{-}+\ldots+w^{(N)}_{-}~,\label{chemN}
 }
where $w^{(j)}_{K}$ and $w^{(j)}_{-}$ are the kaon and negatively charged
hadron  identities of the $j$$^{th}$ particle.
The kaon $w_{K}^{(j)}$ and negatively charged hadron  $w_{-}^{(j)}$
identities are defined as: $w_K^{(j)}=1$ and $w_{-}^{(j)}=0$ if the $j$$^{th}$
particle is a $K^+$,  $w_K^{(j)}=1$ and $w_{-}^{(j)}=1$ if the $j$$^{th}$
particle is a $K^-$, and $w_K^{(j)}=0$ and $w_{-}^{(j)}=1$ if the
$j$$^{th}$ particle is a negative hadron but not a $K^-$.

For the single-particle quantities,
\eq{\label{chem1-single}
\alpha~=~ w_{K}~,~~~~~ \beta ~=~ w_{-}~,
}
one obtains:
\eq{\label{w-k}
& \overline{w_K}=\overline{w_K^2}=\frac{\langle K^+\rangle
+\langle
K^-\rangle }{\langle N\rangle}~\equiv~k_+ +k_-\equiv~k~,\\
& \overline{w_-} =\overline{w_-^2}=\frac{\langle
H^-\rangle}{\langle N\rangle}~\equiv~h_-~,~~~~~
\overline{w_K\,w_-} =\frac{\langle K^-\rangle}{\langle N\rangle}=
k_- ~,\label{w-N}
}
where $N=K^++H^-$. Then from Eqs.~(\ref{w-k}, \ref{w-N}) it follows:
\eq{\label{om-kNNN}
& \omega[w_K]~\equiv~
\frac{\overline{w_K^2}~-~\overline{w_K}^2}{\overline{w_K}}~=~
1~-~k~, ~~~~~~\omega[w_-]~\equiv~
\frac{\overline{w_-^2}~-~\overline{w_-}^2}{\overline{w_-}}~=~
1- h_-~,\\
&\overline{w_Kw_-}~-~\overline{w_K}\,\cdot\,\overline{w_-}~=~k_-~-~~k\cdot
h_-~.\label{K-minus}
}
Therefore, for Eqs.~(\ref{C-D},\ref{C-S}) one finds:
\eq{\label{CD-KN}
C_{\Delta}&~=~
\langle N \rangle \cdot \left(h_- -k\right)~
=~\langle H^-\rangle ~-~\langle K\rangle ~, \\
C_{\Sigma}&~=~
\langle N\rangle \cdot (h_- +k -2 k_-) ~=~ \langle H^- \rangle
~+~\langle K\rangle ~-~ 2\,\langle K^-\rangle ~. \label{CS-KN}
}
Thus in this case the normalization factors depend
on $\langle K \rangle$ and $\langle H^-\rangle$, and
in addition on $\langle K^-\rangle$.

\subsection{The most general case}\label{examp-comp}
The most general case, which up to now was not considered in the literature,
concerns relative fluctuations of two motional extensive
quantities, e.g., energy of charged kaons $E_K$ and transverse momentum
of all negatively charged hadrons $P_T^-$. These two sets of particles
are partly overlapping.
This example corresponds  to the following choice of the extensive
state quantities $A$ and $B$:
\eq{\label{compK}
& A \equiv E_K~=~w^{(1)}_K \epsilon^{(1)}
                +w^{(2)}_K \epsilon^{(2)}
                +\ldots+w^{(N)}_K \epsilon^{(N)} ~,
\\
& B \equiv P_T^-~=~w^{(1)}_- p_t^{(1)}
                +w^{(2)}_- p_t^{(2)}
                +\ldots+w^{(N)}_- p_t^{(N)}~,\label{compN}
}
where $w^{(j)}_K$ and $w^{(j)}_-$ are the kaon and negatively
charged hadron identities of $j$$^{th}$ particle, and $\epsilon^{(j)}$
and $p_t^{(j)}$ are its energy and transverse momentum. Note that
for $\epsilon=p_t=1$ Eqs.~(\ref{compK}) and (\ref{compN}) are
reduced to \mbox{Eqs.~(\ref{chemK}, \ref{chemN})}, respectively.

For the single-particle quantities,
\eq{\label{comp_single}
\alpha~=~ w_K \,\epsilon ~,~~~~~ \beta~=~ w_-\, p_t ~,
}
one obtains:
\eq{\label{wK-com}
& \overline{\alpha}= k\cdot\overline{\epsilon} ~,~~~~
\overline{\alpha^2}=k\cdot\overline{\epsilon^2} ~,~~~~
\overline{\beta} =h_-\cdot \overline{p_t}
~,~~~~~\overline{\beta^2} =h_-\cdot \overline{p_t^2}~,~~~~
 \overline{\alpha \beta}~ =~ k_-\cdot\overline{\epsilon\, p_t} ~,
%
}
where ($n=1,\,2$)
\eq{\label{pt-ep}
& \overline{\epsilon^n} ~=~\int
d\epsilon\,\epsilon^n\,f_K(\epsilon)~,
~~~~\overline{p_t^n}~=~\int dp_t\,p_t^n\, f(p_t)~,\\
& \overline{\epsilon\,p_t}~=~\int d^3p\,\sqrt{{\bf
p}^2+m_K^2}\,p_t\, f_{K^-}({\bf p})~.\label{ptep-cor}
}
In order
to calculate the  averages (\ref{pt-ep}, \ref{ptep-cor}) one needs
to know the single-particle $\epsilon$-distribution for kaons
$f_K(\epsilon)$, the $p_t$-distribution for negatively charged hadrons
$f(p_t)$, and the ${\bf p}$-distribution for $K^-$.
Then from Eq.~(\ref{wK-com}) follows:
\eq{\label{om-kNN}
&
\omega[\alpha]
~=~ \omega[\epsilon]~+~(1-k)\cdot \overline{\epsilon}~,~~~~
\omega[\beta]
~=~
\omega[p_t]~+~(1-h_-)\cdot \overline{p_t}~,\\
&\overline{\alpha\beta}-\overline{\alpha}\cdot
\overline{\beta}~=~k_-\cdot
\overline{\epsilon\,p_t}~-~k\,h_-\,\overline{\epsilon}\cdot\overline{p_t}~.
 }
Finally one finds for the normalisation factors:
\eq{\label{CD-comp}
C_{\Delta}&~=~ \langle P_T^-\rangle \cdot \Big[\,
\omega[\epsilon]~+~\overline{p_t}\,\Big]~-~ \langle E_K\rangle
\cdot \Big[\,\omega[p_t]~+~\overline{\epsilon}\,\Big]
~, \\
C_{\Sigma}&~=~\langle P_T^-\rangle \cdot \Big[\,
\omega[\epsilon]~+~\overline{p_t}\,\Big]~+~ \langle E_K\rangle
\cdot \Big[\,\omega[p_t]~+~\overline{\epsilon}\,\Big]~-~2\,\langle
K^-\rangle\cdot \overline{\epsilon\, p_t}  ~. \label{CS-comp}
}
For the special case $\epsilon=p_t=1$ one gets:
\eq{
\overline{\epsilon}=\overline{\epsilon^2}=
\overline{p_t}=\overline{p^2_t}=\overline{\epsilon\,p_t}=1~,
}
leading to:
\eq{
\omega[\epsilon]=\omega[p_t]=0~,~~~~\langle E_K\rangle \rightarrow
\langle K\rangle~,~~~~\langle P_T\rangle\rightarrow \langle
H^-\rangle~,
}
and Eqs.~(\ref{CD-comp}, \ref{CS-comp}) reduce to
Eqs.~(\ref{CD-KN}, \ref{CS-KN}), respectively.

\newpage

\end{document}